\documentclass{article}
\usepackage{spconf}
\usepackage{color}
\usepackage{theorem}
\usepackage{times,amsmath,epsfig}
\usepackage{amssymb}
\usepackage{subfigure}
\usepackage{cite}

\usepackage{algorithmic}
\usepackage{algorithm}
\usepackage{needspace}

\usepackage{tikz}
\usetikzlibrary{shapes,arrows}
\input{mysymbol.sty}
\tikzstyle{phantom vertex} = [ ellipse, 
                               anchor = center, 
                               minimum height = 1*\unit, 
                               minimum width  = 1*\unit,
                               inner sep=0pt,
                               anchor=center]
\tikzstyle{red vertex}   = [black, fill = red!20,   phantom vertex, draw]
\tikzstyle{black vertex} = [black, fill = black!20, phantom vertex, draw]
\tikzstyle{blue vertex}  = [black, fill = blue!20,  phantom vertex, draw]
\tikzstyle{green vertex} = [black, fill = green!20,  phantom vertex, draw]
\tikzstyle{yellow vertex} = [black, fill = yellow!20,  phantom vertex, draw]
\tikzstyle{cyan vertex} = [black, fill = cyan!20,  phantom vertex, draw]
\tikzstyle{vertex}       = [draw, phantom vertex]

\tikzstyle{point} = [ellipse, inner sep=0pt, draw, fill=white, anchor = center,
                     minimum height = 0.05*\unit, minimum width  = 0.05*\unit]

\newcommand{\QED}{\hfill\ensuremath{\blacksquare}}

\newtheorem{myproposition}{\bf Proposition}

\newenvironment{myproofsketch}[1][$\!\!$]{{\noindent\bf Proof (sketch) #1: }}
{\hfill\QED\medskip}

\title{Network Topology Identification from Spectral Templates}
\name{Santiago Segarra$^{\dag}$, Antonio G. Marques$^{\ddagger}$, Gonzalo Mateos$^{*}$ and Alejandro Ribeiro$^{\dag}$\thanks{Work in this paper is supported by the NSF award  CCF-1217963 and Spanish MINECO TEC2013-41604-R. 
		}}
\address{$^{\dag}$Dept. of ESE, University of Pennsylvania, Philadelphia, PA, USA\\
	$^{\ddag}$Dept. of TSC, King Juan Carlos University, Madrid, Spain\\
	$^{*}$Dept. of ECE, University of Rochester, Rochester, NY, USA}

\begin{document}
\ninept
\maketitle
\begin{abstract}
Network topology inference is a cornerstone problem in statistical analyses of complex systems.
In this context, the fresh look advocated here permeates benefits from convex optimization and graph signal processing, 
to identify the so-termed graph shift operator (encoding the network topology) given only the eigenvectors of the shift. These \emph{spectral templates} can be obtained, for example, from principal component analysis of a set of graph signals defined on the particular network. 
The novel idea is to find a graph shift that while being consistent with the provided spectral information, it endows the network structure with certain desired properties such as sparsity. The focus is on developing efficient recovery algorithms along with identifiability conditions for two particular shifts, the adjacency matrix and the normalized graph Laplacian. 
Application domains include network topology identification from steady-state signals generated by a diffusion process, and design of a graph filter that facilitates the distributed implementation of a prescribed linear network operator. Numerical tests
showcase the effectiveness of the proposed algorithms in recovering synthetic and structural brain networks.
\end{abstract}
%

\section{Introduction}\label{S:Introduction}
\vspace{-0.05in}

Advancing a holistic theory of networks necessitates fundamental breakthorughs in modeling,
identification, and controllability of distributed network processes -- often conceptualized as \emph{signals defined
on the vertices of a graph}~\cite{barrat2012book,kolaczyk2009book}.
Under the assumption that the signal properties are related to the topology of the graph where they are supported, the goal of graph signal processing (GSP) is to develop algorithms that fruitfully leverage this relational structure~\cite{EmergingFieldGSP,SandryMouraSPG_TSP13}.
Instrumental to that end is the so-termed graph-shift operator (GSO)~\cite{SandryMouraSPG_TSP13}, a matrix capturing the graph's local topology and whose eigenbasis is central to defining graph Fourier transforms~\cite{SandryMouraSPG_TSP14Freq}. Most GSP works assume that the GSO (hence the graph) is known, and then analyze how the algebraic and spectral characteristics of the GSO impact the properties of the signals and filters defined on such a graph. Here instead we take the reverse path and investigate how to use information available from graph signals and filters to infer the underlying graph topology; see also~\cite{DongLaplacianLearning,MeiGraphStructure}. In a nutshell, we follow a two-step approach whereby we first leverage results from GSP theory to identify the GSO's eigenbasis using the available information, and then rely on these \emph{spectral templates} to recover the GSO itself.


Network topology inference from a set of (graph-signal) observations is a prominent problem in Network Science~\cite{kolaczyk2009book,sporns2012book}. Since networks encode similarities between nodes, most approaches construct graphs whose edge weights correspond to the correlation, or coherence between signal profiles at incident nodes. 
This approach is not without merit and widely used in practice, but it exhibits several drawbacks, the main one being that links are formed taking into account only pairwise interactions, ignoring that the observed correlations can be due to latent network effects. Acknowledging these limitations, alternative methods rely on partial correlations~\cite{GLasso2008,kolaczyk2009book}, Gaussian graphical models~\cite{slawski2015estimation,meinshausen06}, or, Granger causality~\cite{Brovelli04Granger,sporns2012book}.  Differently, recent GSP-based network inference frameworks postulate that the network exists as a latent underlying structure, and that observations are generated as a result of a network process defined in such graph. For instance, network structure is estimated in~\cite{MeiGraphStructure}  to unveil unknown relations among nodal time series adhering to an autoregressive model involving graph-filter dynamics. A factor analysis-based approach was put forth in~\cite{DongLaplacianLearning} to infer graph Laplacians, seeking that input graph signals are smooth over the learned topologies. Different from~\cite{DongLaplacianLearning,MeiGraphStructure} that operate on the graph domain, the goal here is to identify graphs that endow the given observations with desired spectral (frequency-domain) characteristics.

After surveying the required GSP background in Section \ref{S:GSP_101}, we formulate the GSO identification problem given spectral templates (Section \ref{S:ShiftInfFullEigen}). The novel idea is to search among all feasible networks for the one that endows the resulting graph-signal transforms with prescribed spectral properties, while the inferred graph also exhibits desirable structural characteristics.  Optimization problems are formulated to recover two particular GSOs, namely the adjacency matrix (Section \ref{Ss:Adjacency}) and the normalized graph Laplacian (Section \ref{Ss:Laplacian}). Conditions under which the feasible set reduces to a singleton are derived, and convex relaxations leading to computationally-efficient algorithms are proposed for the general case. Section \ref{S:ObtainingSpecTemplates} outlines two pragmatic scenarios where acquisition of the GSO eigenbasis is well motivated, making the case that setups where the GSO is unknown but its spectral templates are available can be more the rule than the exception. Computer simulations highlight the effectiveness of the proposed algorithms in identifying both synthetic and real-world networks (Section \ref{S:Simulations}).  Concluding remarks are given in Section \ref{S:Conclusions}. 

\section{Graph signals and graph filters}\label{S:GSP_101}
\vspace{-0.05in}

Here we formally introduce notation and terminology as well as different GSP tools that will be used throughout the paper. 

\noindent\textbf{Graphs.} Let $\ccalG$ denote an undirected graph with a set of nodes $\mathcal{N}$ (with cardinality $N$) and a set of links $\ccalE$, such that if node $i$ is connected to $j$, then both $(i,j)$ and $(j,i)$ belong to $\ccalE$. The set $\mathcal{N}_i:=\{j\:|(j,i)\in\mathcal{E}\}$ stands for the neighborhood of $i$. 
For any given $\mathcal{G}$ we define the adjacency matrix $\bbA\in\mathbb{R}^{N\times N}$ as a sparse matrix with non-zero elements $A_{ij}=A_{ji}$ if and only if $(i,j)\in\ccalE$. The values of $A_{ji}$ can be binary, or real in the weighted case to capture the strength of the connection from $i$ to $j$. Matrix $\bbA$ can be used to define the degree vector $\bbd:=\bbA\bbone$ and the degree matrix $\bbD:=\diag(\bbd)$. Moreover, the normalized Laplacian matrix is $\bbL:=\bbI - \bbD^{-1/2} \bbA\bbD^{-1/2}\in\mathbb{R}^{N\times N}$, which is positive semidefinite and has eigenvector $\sqrt{\bbd}:=\bbD^{1/2}\mathbf{1}$ with associated eigenvalue zero. 

\noindent\textbf{Graph signals and shift operator.} 
Graph signals defined on the nodes of $\ccalG$ are functions $f : \mathcal{N} \to \mathbb{R}$, equivalently represented as vectors $\mathbf{x}=[x_1,...,x_N]^T \in  \mathbb{R}^N$, where $x_i$ denotes the signal value at node $i$. Since $\bbx$ does not account explicitly for the structure of the graph where the signal is defined, $\mathcal{G}$ is endowed with the so-called GSO $\mathbf{S}$ \cite{SandryMouraSPG_TSP13,SandryMouraSPG_TSP14Freq}. The shift $\mathbf{S}\in\mathbb{R}^{N\times N}$ is a matrix whose entry $S_{ij}$ can be non-zero only if $i=j$ or if $(i,j)\in\mathcal{E}$. The sparsity pattern of  $\bbS$ captures the local structure of $\ccalG$, but we make no specific assumptions on the values of its non-zero entries. The shift $\mathbf{S}$ can also be understood as a linear transformation that can be computed locally at the nodes of the graph. More rigorously, if $\mathbf{y}$ is defined as $\mathbf{y}=\mathbf{S}\mathbf{x}$, then node $i$ can compute $y_i$ provided that it has access to the value of $x_j$ at $j\in \mathcal{N}_i$.
Typical choices for $\mathbf{S}$ are the adjacency matrix $\bbA$ \cite{SandryMouraSPG_TSP13,SandryMouraSPG_TSP14Freq}, the (normalized) Laplacian $\bbL$~\cite{EmergingFieldGSP}, and their respective generalizations \cite{godsil2001algebraic}. We assume henceforth that $\bbS$ is symmetric, so that $\bbS=\bbV\bbLambda\bbV^{T}$ with $\bbLambda\in\mathbb{R}^{N\times N}$ being diagonal, but our results hold for any normal GSO.

\noindent\textbf{Graph filters.} The shift $\bbS$ can be used to define linear, shift-invariant graph-signal \emph{operators} of the form
\begin{eqnarray}\label{E:Filter_input_output_time}
	&\mathbf{H}:=\sum_{l=0}^{L-1}h_l \mathbf{S}^l&
\end{eqnarray}
which are called \emph{graph filters} \cite{SandryMouraSPG_TSP13}. For a given input $\bbx$, the output of the filter is simply $\bby=\bbH\bbx$. The coefficients of the filter are collected into $\mathbf{h}:=[h_0,\ldots,h_{L-1}]^T$, with $L-1$ denoting the filter degree. Graph filters are of particular interest because they represent linear transformations that can be implemented locally~\cite{EUSIPCO_our_interp_2015,ssamar_distfilters_allerton15}.

\noindent\textbf{Frequency domain representation.} Leveraging the spectral decomposition of $\bbS$, graph filters and signals can be represented in the frequency domain.
To be precise, let us use the eigenvectors of $\bbS$ to define the $N\times N$ matrix $\bbU:=\bbV^T$, and the eigenvalues of $\bbS$ to define the $N\times L$ Vandermonde matrix $\bbPsi$, where $\Psi_{ij}:=(\Lambda_{ii})^{j-1}$.
Using these conventions, the frequency representations of a \emph{signal} $\bbx$ and of a \emph{filter} $\bbh$ are defined as $\widehat{\bbx}:=\bbU\bbx$ and $\widehat{\bbh}:=\bbPsi\bbh$, respectively~\cite{SandryMouraSPG_TSP14Freq}. 
Exploiting such representations, the filter's output $\bby\!=\!\bbH\bbx$ in the frequency domain is ($\odot$ denotes Hadamard product)
\begin{equation}\label{E:Filter_input_output_freq}
	\widehat{\bby}=\diag\big(\bbPsi\bbh\big)\bbU \bbx=\diag\big(\widehat{\bbh}\big)\widehat{\bbx}=\widehat{\bbh}\odot\widehat{\bbx}.
\end{equation}
This identity is the counterpart of the convolution theorem for temporal signals
; see e.g. \cite{SSGMAMAR_camsap15} for a derivation.
Interestingly, while in the time domain $\bbU=\boldsymbol{\Psi}$ and they correspond to the Discrete Fourier Transform (DFT) matrix, this is not true for general graphs.

\noindent\textbf{Network diffusion processes.} Graph filters can be used to model network diffusion processes. 
Specifically, the signal at node $i$ during the step $(l+1)$ of a linear diffusion process can be written as
\begin{eqnarray}
\label{E:local_diffusion_singlenode}
&x_i^{(l+1)}=\alpha_{i,i} x_{i}^{(l)} + \sum_{j\in \ccalN_i} \alpha_{i,j} x_{j}^{(l)}&
\end{eqnarray}
where 
$\alpha_{i,j}$ are the diffusion coefficients; see e.g.,~\cite{SSGMAMAR_camsap15}. Leveraging the GSP framework, \eqref{E:local_diffusion_singlenode} implies that the graph signal $\bbx^{(l+1)}=\bbS\bbx^{(l)}$ at iteration $l+1$ is the shifted version of $\bbx^{(l)}$, for a shift $\bbS$ with entries  $S_{ij}=\alpha_{i,j}$ if either $i=j$ or $(i,j) \in \ccalE$, and $S_{ij}=0$ otherwise.
For instance, if we set $\bbS=\bbI - \beta \bbL$ and let the signal of interest be $\bbx := \bbx^{(\infty)}$, then $\bbx$ solves the heat diffusion equation~\cite{segarra2015reconstruction,segarra2015blindCAMSAP}. However, more complex diffusion dynamics such as $\bbx=\Pi_{l=0}^{\infty}(\bbI-\beta_l \bbS)\bbx^{(0)}$ and $\bbx=\sum_{l=0}^{\infty}\gamma_l \bbx^{(l)}=\sum_{l=0}^{\infty}\gamma_l \bbS^l\bbx^{(0)}$, could also be of interest. 

The Cayley€"-Hamilton theorem guarantees that the aforementioned infinite-horizon processes can be equivalently described by a filter of degree $N$.
Accordingly, several works have recognized that the steady-state signal $\bbx$ generated by a diffusion process can be modeled as the output of a graph filter $\bbH=\sum_{l=0}^{N-1}h_l \bbS^l$ with input (seed) $\bbx^{(0)}$~\cite{segarra2015reconstruction,segarra2015blindCAMSAP}.
This key insight can be used to relate the statistical and spectral properties of $\bbx$ and $\bbH$, which will be leveraged in the ensuing sections to identify $\bbS$ itself. 


\section{Shift inference from spectral templates}\label{S:ShiftInfFullEigen}
\vspace{-0.05in}

Given a set of eigenvectors $\bbV=[\bbv_1,\ldots,\bbv_N]$, also termed \emph{spectral templates}, our goal is to find a graph shift $\bbS$ that is diagonalized by $\bbV$. As detailed in Section \ref{S:ObtainingSpecTemplates}, the knowledge of $\bbV$ can for instance be obtained from the PCA decomposition of signals like the ones generated by the diffusion processes outlined in Section \ref{S:GSP_101}; and whose covariance depends on the structure of $\bbS$. Since the postulated problem has infinitely many solutions, we further impose conditions on $\bbS$ promoting desirable properties such as sparsity, or a priori information on the graph of interest such as non-negative edge weights.
Note that by definition, $\bbS$ encodes the local structure of the graph it represents, thus, its recovery implies a successful identification of the graph topology of interest.

\vspace{-0.05in}
\subsection{Identifying the adjacency matrix}\label{Ss:Adjacency}
\vspace{-0.05in}

Consider the unknown adjacency matrix $\bbS=\bbA$ of an unweighted and undirected graph, and suppose its spectral templates $\bbV$ are given. With $\bblambda=[\lambda_1,...,\lambda_N]^T$ collecting the unknown eigenvalues of $\bbS$ (hence $\bbLambda=\text{diag}(\bblambda)$), we aim at identifying $\bbS$ by solving
\begin{align}
	\label{E:SparseAdj_l00_obj} \min_{ \{\bbS, \bblambda\}} \;\;&\| \bbS\|_{0}\\
	\label{E:SparseAdj_l00_const1}\text{s. to } \;\;&\;\bbS=\textstyle\sum_{k =1}^N \lambda_k\bbv_k \bbv_k^T,\\ \label{E:SparseAdj_l00_const2}&\;S_{ij}\in\{0,1\}, \;\; S_{ii}=0, \;\; \bbS\in\ccalM^N,\;\; \bbS \mathbf{1}\geq \mathbf{1}
\end{align}
where $\ccalM^N$ denotes the set of $N\times N$ real and symmetric matrices. Among the potentially multiple feasible solutions,  the cardinality function $\|\bbS\|_{0}$ in the objective selects as optimum the one that minimizes the number of edges. The role of constraint $\bbS \bbone \geq \bbone$ (entrywise), which reasonably requires each node to have at least one neighbor, is to prevent the trivial solution $\bbS=\bb0$.
As can be seen from \eqref{E:SparseAdj_l00_obj}-\eqref{E:SparseAdj_l00_const2}, when all eigenvectors $\{\bbv_k\}_{k=1}^N$ are given the design of $\bbS$ amounts to finding the $N$ eigenvalues in $\bblambda$. The constraint in \eqref{E:SparseAdj_l00_const1} encodes the definition of a general graph shift $\bbS = \bbV \bbLambda \bbV^T$, while those in \eqref{E:SparseAdj_l00_const2} incorporate the supplementary conditions implied by the fact that $\bbS$ is an adjacency matrix.
Additional information can be incorporated in \eqref{E:SparseAdj_l00_const2}, such as a priori knowledge on particular entries of $\bbS$ or node degrees. Alternatively, the feasible set can be enlarged to accommodate for adjacency matrices with negative entries.

An interesting property of the proposed optimization is that the feasible set described by \eqref{E:SparseAdj_l00_const1}-\eqref{E:SparseAdj_l00_const2} is generally small. To be more precise, we define the matrix $\bbW  := \bbV\odot\bbV \in \reals_+^{N\times N}$ and denote by $Q$ the number of singular values of $\bbW$ that are zero. 

%
\begin{myproposition}\label{P:Feasibility_Adjacency}
	Assume that \eqref{E:SparseAdj_l00_obj}-\eqref{E:SparseAdj_l00_const2} is feasible, then it holds that:\\
	a) The nullspace of $\bbW$ has dimension at least one, so that $Q\geq 1$.\\
	b) If $Q=1$, the feasible set given by \eqref{E:SparseAdj_l00_const1}-\eqref{E:SparseAdj_l00_const2} is a singleton. 
\end{myproposition}
%
\vspace{-0.05in}
\noindent {\bf Proof (sketch):} The key of the proof resides in noting that the columns of $\bbW$ correspond to the diagonal entries of $\bbv_k \bbv_k^T$ for $k = 1, \ldots, N$. Then, by combining \eqref{E:SparseAdj_l00_const1} and \eqref{E:SparseAdj_l00_const2} it follows that $\bbW \bblambda = \mathbf{0}$ for all feasible $\bblambda$, thus, feasibility implies $Q\geq 1$. When $Q=1$, $\bblambda$ (and hence $\bbS$) is unique up to a scaling factor. Therefore using the fact that $\bbS$ is binary, statement b) follows. {\hfill\QED\medskip}


\noindent \textbf{Relaxation and algorithmic discussion.}
The two sources of non-convexity that render the solution to \eqref{E:SparseAdj_l00_obj}-\eqref{E:SparseAdj_l00_const2} challenging are the presence of the $\ell_0$ norm in the objective and the binary constraints $S_{ij}\in\{0,1\}$. We relax the former to an iteratively re-weighted $\ell_1$ norm and the latter by replacing $\{0,1\}$ with its convex hull $[0,1]$. Specifically, with $p$ denoting an iteration index, we aim to solve a sequence $p=1,...,P$ of $\ell_1$-norm penalized problems
\begin{align}
	\label{E:SparseAdj_wl11_obj} \min_{ \{\bbS, \bblambda\}} \;\;& { \textstyle\sum_{i=1}^N\sum_{j=1}^N} \omega_{ij}(p)|S_{ij}| \\
	\label{E:SparseAdj_wl11_const1}\text{s. to } \;\;&\;\bbS=\textstyle\sum_{k =1}^N \lambda_k\bbv_k \bbv_k^T,\\ \label{E:SparseAdj_wl11_const2}&\;S_{ij}\in[0,1], \;\; S_{ii}=0, \;\; \bbS\in\ccalM^N,\;\; \bbS \mathbf{1}\geq \mathbf{1}
\end{align}
with weights $\omega_{ij}(p):= \left(|S_{ij}(p-1)|+\delta\right)^{-1}$ where $\delta$ is a small and positive constant. Intuitively, the goal of the re-weighted scheme in \eqref{E:SparseAdj_wl11_obj} is that if the value of $|S_{ij}(p-1)|$ is small, in the next iteration the penalization weight $\omega_{ij}(p)$ is large, promoting further shrinkage of $S_{ij}$ towards zero; see, e.g., \cite{candes_l0_surrogate} for technical details. The recovery performance of the proposed scheme is analyzed in Section~\ref{S:Simulations}.

Building on Prop.~\ref{P:Feasibility_Adjacency}, whenever $Q = 1$ the solutions of \eqref{E:SparseAdj_wl11_obj}-\eqref{E:SparseAdj_wl11_const2} and \eqref{E:SparseAdj_l00_obj}-\eqref{E:SparseAdj_l00_const2} coincide up to a scaling factor. 
\begin{myproposition}\label{P:Identifiability_Adjacency_Norm1}
	Let $\bbS_0^*$ and $\bbS_1^*$ denote solutions to \eqref{E:SparseAdj_l00_obj}-\eqref{E:SparseAdj_l00_const2} and \eqref{E:SparseAdj_wl11_obj}-\eqref{E:SparseAdj_wl11_const2}, respectively. If $Q=1$, then $\bbS_1^*=d_{\min}^{-1} \bbS_0^*$, with $d_{\min}$ being the minimum node degree in the graph to recover.    
\end{myproposition}
\vspace{-0.05in}
\begin{myproofsketch}
	Denoting by $\bblambda_0^*$ the unique solution to \eqref{E:SparseAdj_l00_obj}-\eqref{E:SparseAdj_l00_const2} (cf.~Prop.~\ref{P:Feasibility_Adjacency}), the solution to \eqref{E:SparseAdj_wl11_obj}-\eqref{E:SparseAdj_wl11_const2} is $\bblambda_1^* =  \alpha \bblambda_0^*$  with $\alpha$ being the smallest positive number satisfying $\bbS \mathbf{1}\geq \mathbf{1}$, which is $\alpha=d_{\min}^{-1}$.
\end{myproofsketch}


\vspace{-0.05in}
\subsection{Identifying the Laplacian matrix and generalized shifts}\label{Ss:Laplacian}
\vspace{-0.05in}

Consider the unknown normalized Laplacian matrix $\bbS=\bbL$ of an unweighted and undirected graph. With $\ccalM_{+}^N$ denoting the conical set of positive semidefinite $N\times N$ matrices, the problem to solve now is [cf. \eqref{E:SparseAdj_l00_obj}-\eqref{E:SparseAdj_l00_const2}]
\begin{align}
	\label{E:SparseLap_l00_obj} \min_{ \{\bbS, \bblambda\}} \;\;&\| \bbS\|_{0}\\
	\label{E:SparseLap_l00_const1}\text{s. to } \;\;&\;\bbS=\textstyle\sum_{k =1}^N \lambda_k\bbv_k \bbv_k^T,\;\;\;\; \bbS\! \in \! \ccalM_{+}^N, \;\;\;\; \lambda_1=0,\\ 
	\label{E:SparseLap_l00_const2}&\;\;\,\,\,\, S_{ij} \in [-1, 0]  \,\, \text{for} \,\, i\neq j, \qquad S_{ii}=1 \,\, \forall \,\, i.
\end{align}
In the above formulation we assumed that $\bbv_1$ corresponds to the eigenvector $\sqrt{\bbd}$ so that we can leverage the structure of normalized Laplacians and enforce $\lambda_1=0$. This, in turn, prevents the trivial solution $\bbS=\bbI$. Note also that the eigenvector $\sqrt{\bbd}$ can be easily identified from the $N$ eigenvectors in $\bbV$ since it is the only one whose entries have all the same sign \cite{biyikougu2007laplacian}. As was the case for \eqref{E:SparseAdj_l00_obj}-\eqref{E:SparseAdj_l00_const2}, if more than one feasible solution exists, the identified Laplacian corresponds to the topology that minimizes the number of edges. 

Interestingly, problem \eqref{E:SparseLap_l00_obj}-\eqref{E:SparseLap_l00_const2} also has a small feasible set. To state this formally, consider the matrix $\tilde{\bbV}=[\bbone, \bbv_2, \bbv_3, \ldots, \bbv_N ]$, define $\tilde{\bbW}:=\tilde{\bbV}\odot\tilde{\bbV}$ and denote by $\tilde{Q}$ the number of singular values of $\tilde{\bbW}$ that are zero. Then the following result, which is the counterpart of Prop.~\ref{P:Feasibility_Adjacency} for Laplacian identification, holds.
\begin{myproposition}\label{P:Feasibility_normLaplacian}
	Assume that \eqref{E:SparseLap_l00_obj}-\eqref{E:SparseLap_l00_const2} is feasible, then it holds that:\\
	a) The nullspace of $\tilde{\bbW}$ has at least dimension one, so that $\tilde{Q}\geq 1$.\\
	b) If $\tilde{Q}=1$, the feasible set given by \eqref{E:SparseLap_l00_const1}-\eqref{E:SparseLap_l00_const2} is a singleton.
\end{myproposition}

The result follows using arguments similar to those in the proof of Prop.~\ref{P:Feasibility_Adjacency}. Regarding the design of efficient algorithms to solve \eqref{E:SparseLap_l00_obj}-\eqref{E:SparseLap_l00_const2}, since \eqref{E:SparseLap_l00_obj} is the only source of non-convexity, the iteratively re-weighted scheme in \eqref{E:SparseAdj_wl11_obj} can also be applied here. Details are omitted due to space limitations.

Modifications to the proposed formulations can be made to accommodate for shifts that exhibit properties other than sparsity. For example, one could go after Laplacians with good mixing conditions, related to having a large second smallest eigenvalue \cite{chung1997spectral}. This can be accomplished by introducing the optimization variable $\lambda_{\min}$, adding the constraints $\lambda_{\min}\leq\lambda_k$ for all $k\geq 2$ and augmenting the cost with the regularization $-\eta \lambda_{\min}$, with $\eta$ being a tuning constant.

The identification of graph shifts different from the adjacency and the normalized Laplacian can be of interest, including the combinatorial Laplacian $\bbL=\bbD - \bbA$ and the random walk Laplacian $\bbL=\bbD^{-1}\bbA$. These would require minor modifications in the constraints \eqref{E:SparseLap_l00_const1}-\eqref{E:SparseLap_l00_const2}. For the particular case of the combinatorial Laplacian, results in Prop.~\ref{P:Feasibility_normLaplacian} can be extended directly provided that the degree vector $\bbd$, which contains the same information than the first eigenvector of the normalized Laplacian, is known. Specifically, the constraints $S_{ii}=1$ in \eqref{E:SparseLap_l00_const2} need to be replaced with $S_{ii}=d_i$ and matrix $\tilde{\bbV}$, which is used in the definition of $\tilde{\bbW}=\tilde{\bbV}\odot\tilde{\bbV}$, has to be redefined as $\tilde{\bbV}=[\bbd, \bbv_2, \bbv_3, \ldots, \bbv_N ]$. 


\begin{figure*}[t]
	\centering
	\input{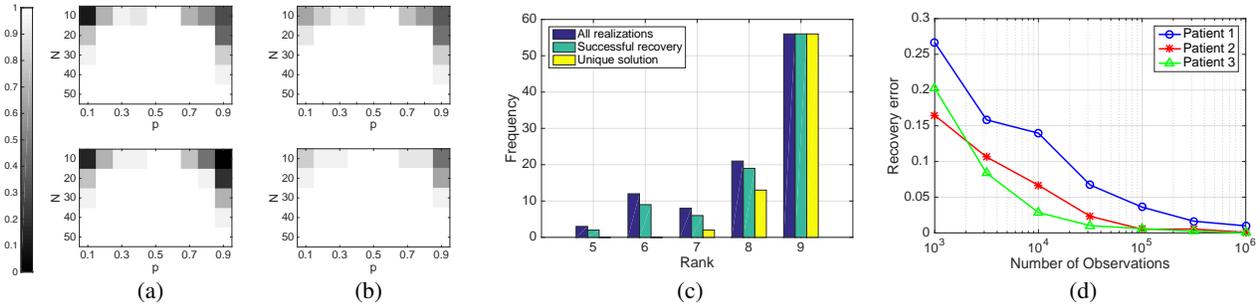}	
	\vspace{-0.175in}
	\caption{(a) Proportion of topology identification problems with unique solution for Erd\H{o}s-R\'enyi graphs as a function of their size $N$ and the probability of edge appearance $p$ for adjacency (top) and normalized Laplacian (bottom) matrices. (b) Recovery rate for the same set of graphs used in (a) when implementing the iteratively re-weighted approach. (c) Histogram of the rank of matrix $\tilde{\bbW}$ for $N=10$ and $p=0.2$. (d) Brain graph recovery error for three patients as a function of the number of signals observed in the estimation of the spectral templates. }
		\vspace{-0.15in}
	\label{F:num_exp}
\end{figure*}

\section{Obtaining the graph spectral templates}\label{S:ObtainingSpecTemplates}
\vspace{-0.05in}

The methods discussed in Section~\ref{S:ShiftInfFullEigen} are relevant in practice when we have access to the eigenvectors (i.e., the spectral templates) of $\bbS$, but not to the GSO itself. A couple of scenarios where that is indeed the case are presented next.

\vspace{0.05in}
\noindent\textbf{Signals generated through diffusion processes.}
Recall that graph filters can be used to model diffusion processes that depend on the topology of the graph (cf.~Section~\ref{S:GSP_101}). More specifically, a diffused or steady-state signal $\bbx$ can written as $\bbx=\bbH \bbx^{(0)}$ where $\bbx^{(0)}$ is a seed signal and $\bbH=\sum_{l=0}^{N-1}h_l \bbS^l$ is a graph filter.
Under the assumption that $\bbx^{(0)}$ is white (with identity covariance matrix) and zero mean, the covariance matrix of the output $\bbC_\bbx:=\E{\bbx\bbx^T}$ is given by
\vspace{-0.03in}
\begin{align}\label{E:cov_output_filter}
\hspace{-0.28cm}\bbC_\bbx=\bbH \E{\bbx^{(0)}({\bbx^{(0)}})^T}\bbH^T=\bbH\bbH^T\stackrel{\star}{=}\bbV\diag(|\widehat{\bbh}|^2)\bbV^T,
\end{align}
\vspace{-0.03in}
where for equality $\stackrel{\star}{=}$ we leveraged the frequency interpretation of a graph filter. Expression \eqref{E:cov_output_filter} reveals that the eigenvectors of the covariance matrix $\bbC_\bbx$ and those of $\bbS$ are the same. Thus, if $\bbC_\bbx$ is known, the spectral templates $\bbV$ can be readily obtained. More commonly, if $\bbC_\bbx$ is unknown but we have access to a set of $M$ diffused signals $\{\bbx_m\}_{m=1}^M$, we may approximate $\bbC_\bbx$ with the sample covariance $\hat{\bbC}_\bbx=1/M\sum_{m=1}^M \bbx_m \bbx_m^T$ and estimate the eigenvectors of $\bbS$. Applying the methods in Section~\ref{S:ShiftInfFullEigen}, we can then use the estimated eigenvectors $\hat{\bbV}$ to recover $\bbS$ as illustrated in Section~\ref{S:Simulations}. Detailed analysis on the sensibility of the topology identification performance to the errors in estimating $\hat{\bbV}$ is left as future work.

\vspace{0.05in}
\noindent\textbf{Implementation of linear network operators.}
Matrices $\bbB \in \reals^{N \times N}$ can be viewed as linear network operators since they are linear maps between graph signals. Conceivably, one might be interested in representing (or approximating) a pre-specified operator $\bbB$ as a graph filter $\bbH$, to facilitate its distributed implementation in the network. This problem is thoroughly investigated in \cite{segarra2015graphfilteringTSP15}, and one of the conditions derived for perfect implementation states that all the eigenvectors of $\bbB$ and $\bbS$ must coincide \cite[Prop. 1]{segarra2015graphfilteringTSP15}. Consequently, we can use the eigenvectors of a desired $\bbB$ as spectral templates to generate a graph shift $\bbS$. This way, $\bbB$ can be expressed as a graph filter and implemented via local agent interactions.
%

\vspace{-0.05in}
\section{Numerical experiments}\label{S:Simulations}
\vspace{-0.05in}

We illustrate the performance of the topology identification algorithm by solving \eqref{E:SparseAdj_wl11_obj}-\eqref{E:SparseAdj_wl11_const2}, as well as its counterpart for Laplacian identification, for different synthetic and real-world graphs. 

\vspace{0.05in}
\noindent\textbf{Random graphs.}
Consider Erd\H{o}s-R\'enyi graphs \cite{bollobas1998random} of varying size $N \in \{10, 20, \dots, 50\}$ and different edge-formation probabilities $p \in \{0.1, 0.2, \ldots, 0.9\}$. For each combination of $N$ and $p$ we generate 100 graphs and try to recover their adjacency $\bbA$ and normalized Laplacian $\bbL$ matrices from the corresponding spectral templates $\bbV$. In Fig.~\ref{F:num_exp}(a) we plot the proportion of instances where the corresponding optimization problems -- \eqref{E:SparseAdj_l00_obj}-\eqref{E:SparseAdj_l00_const2} for $\bbA$ and \eqref{E:SparseLap_l00_obj}-\eqref{E:SparseLap_l00_const2} for $\bbL$ -- have unique solutions. Notice that multiple solutions are more frequent when the expected number of neighbors of a given node is close to either $1$ or $N$. For intermediate values of $p$, the rank of both $\bbW$ and $\tilde{\bbW}$ is typically $N-1$, guaranteeing that the solution to our optimization is unique (cf. Props.~\ref{P:Feasibility_Adjacency} and~\ref{P:Feasibility_normLaplacian}). Using the same set of graphs that those in Fig.~\ref{F:num_exp}(a), Fig.~\ref{F:num_exp}(b) shows the recovery rate when solving the iteratively re-weighted problem in \eqref{E:SparseAdj_wl11_obj}-\eqref{E:SparseAdj_wl11_const2} and its counterpart for the Laplacian. As expected, the rates in Fig.~\ref{F:num_exp}(b) dominate those in Fig.~\ref{F:num_exp}(a) since every instance with a unique solution is recovered successfully. Moreover, the improvement in the rates observed in Fig.~\ref{F:num_exp}(b) reflect the effect of the weighted $\ell_1$ norm objective [cf.~\eqref{E:SparseAdj_wl11_obj}] in recovering the true underlying graph.

As indicated by Props.~\ref{P:Feasibility_Adjacency} and~\ref{P:Feasibility_normLaplacian}, the rate of recovery is intimately related to the ranks of $\bbW$ and $\tilde{\bbW}$ for the adjacency and normalized Laplacian case, respectively. Fig.~\ref{F:num_exp}(c) further illustrates this relation via a histogram of the rank of $\tilde{\bbW}$ for the 100 graphs with $N=10$ and $p=0.2$. For more than half of the instances, the rank of $\tilde{\bbW}$ was equal to 9 (blue bar) and, as stated in Prop.~\ref{P:Feasibility_normLaplacian}, for all these graphs the solution was unique (yellow bar) and successfully recovered (cyan bar). We see that, as the rank of $\tilde{\bbW}$ degrades, uniqueness is no longer guaranteed but for most cases the true graph can still be recovered following the iteratively re-weighted scheme proposed. Only in 8 of the cases where $\mathrm{rank}(\tilde{\bbW})<9$ the recovery was not successful, entailing a recovery rate of 0.92, as reported in the corresponding entry ($N=10$, $p=0.2$) of the lower plot in Fig.~\ref{F:num_exp}(b).

\vspace{0.05in}
\noindent\textbf{Brain graphs.}
We consider the identification of unweighted and undirected graphs corresponding to human brains~\cite{hagmann2008mapping}, consisting of $N=66$ nodes or regions of interest (ROIs) and whose edges link ROIs with density of anatomical connections greater than a threshold. The threshold is chosen as the largest one that entails a connected graph. We test the recovery performance for noisy spectral templates $\hat{\bbV}$ obtained from sample covariances of signals generated through diffusion processes (cf.~Section~\ref{S:ObtainingSpecTemplates}). 
Our framework can handle the recovery of $\bbS$ from noisy versions of $\bbV$ by rewriting the equality constraint \eqref{E:SparseAdj_wl11_const1} as an element-wise inequality bounding the absolute difference between $S_{ij}$ and $\textstyle\sum_{k =1}^N \lambda_k V_{ki} V_{kj}$ for all $i,j$. Denoting by $\hat{\bbV}_i$ the noisy spectral templates of patient $i \! \in \! \{1, 2, 3\}$ and by $\hat{\bbA}_i$ the adjacency matrices recovered from them, Fig.~\ref{F:num_exp}(d) plots the recovery error as a function of the number of signals observed in the computation of the sample covariance. The error is quantified as the proportion of edges misidentified, i.e., $\|\bbA_i - \hat{\bbA}_i\|_0 / \| \bbA_i \|_0$, and each point in Fig.~\ref{F:num_exp}(d) is the average across 50 random realizations. 
First notice that for an increasing number of observed signals we see a monotonous decrease in recovery error. For example, when going from $10^4$ to $10^5$ observations the error is (approximately) divided by seven, when averaged across patients. This is reasonable since a larger number of observations gives rise to a more reliable estimate of the covariance matrix entailing a less noisy version of the spectral templates. Notice that, even though the performance increases for all patients with the number of observations, the brain of patient 1 is consistently the hardest to identify. For instance, consider the errors for $10^5$ observations, where for patient 1 we successfully recover $96.4\%$ of the \textit{edges}, whereas for patients 2 and 3 the average recovery rate is $99.5\%$ and $99.4\%$, respectively. This points towards the fact that some graphs are inherently more robust for identification when given noisy spectral templates. A formal analysis of this phenomenon is left as future work.

Traditional methods like graphical lasso \cite{GLasso2008} fail to recover $\bbS$ from the sample covariance of filtered white signals. This occurs because, in general, the filter $\bbH$ introduces conditional dependence between signal values more than one hop apart. Even when focusing on filters of the form $\bbH = h_0 \bbI + h_1 \bbS$, graphical lasso underperforms compared to the method presented. More precisely, based on $10^5$ observations, the recovery error of graphical lasso averaged over 50 realizations and with optimal tuning parameters is 0.303, 0.350, and 0.270 for patients 1, 2, and 3, respectively [cf. Fig.~\ref{F:num_exp}(d)].

\vspace{-0.05in}
\section{Conclusions}\label{S:Conclusions}
\vspace{-0.05in}

We studied the problem of identifying a graph shift operator $\bbS$, i.e., the topology of a graph $\ccalG$ of interest, given its eigenbasis $\bbV$. The focus was on setups where $\bbS$ represents the adjacency or the normalized Laplacian of $\ccalG$. We formulated optimization problems to recover $\bbS$, presented their convex relaxations, and derived conditions under which the relaxed problem is guaranteed to recover the desired shift. To highlight the practical relevance of the proposed schemes we outlined scenarios where $\bbV$ is known but $\bbS$ is not, and carried out numerical tests showing the effectiveness in recovering synthetic and brain graphs, even from imperfect spectral templates.







\bibliographystyle{IEEEbib}
\bibliography{citations}

\end{document}